\documentclass{PoS}
\usepackage{natbib}
\title{Observing the Signatures of the r-Process \\in the Oldest Galactic Stars}

\ShortTitle{Observing the r-Process Signature in the Oldest Stars}

\author{\speaker{Anna Frebel}

        McDonald Observatory, University of Texas at Austin\\
        E-mail: \email{anna@astro.as.utexas.edu}}

      \abstract{The abundance patterns of metal-poor stars provide us
        a wealth of chemical information about various stages of the
        chemical evolution of the Galaxy.  In particular, these stars
        allow us to study the formation and evolution of the elements
        and the involved nucleosynthesis processes. This knowledge is
        invaluable for our understanding of the cosmic chemical
        evolution and the onset of star- and galaxy
        formation. Metal-poor stars are the local equivalent of the
        high-redshift Universe, and thus offer crucial observational
        constraints on a variety of issues regarding the early
        Universe.  This review presents an introduction to metal-poor
        stars and their role as ``cosmic lab'' for the study of
        neutron-capture nucleosynthesis processes, particularly that
        of the r-process. The metal-poor star HE~1523$-$0901 serves as
        an example for this group of objects. It displays in its
        spectrum the strongest overabundance of neutron-capture
        elements associated with the r-process. Heavy neutron-capture
        elements such as Eu, Os, and Ir were measured, as well as the
        radioactive elements Th and U. Abundance of Th and U, in
        conjunction with those of stable elements make possible
        nucleo-chronometry, i.e., the determination of stellar
        ages. HE~1523$-$0901 appears to be $\sim13$\,Gyr old. Age
        uncertainties range from 2 to 5\,Gyr for individual
        chronometers, and are largly due to theoretical uncertainties
        in the initial production ratio of the employed
        chronometers. The decay product of the radioactive elements,
        lead, can be used to constrain r-process calculations. Only
        few such stars are currently known with detected U. These
        objects, however, are crucial for the study of this
        nucleosynthesis process. Once more objects are discovered, and
        assuming an old age for them (infered from their low
        metallicity), stars with measured Th \textit{and} U abundances
        can become stellar age calibrators. This way, ages of stars
        in which only Th is measured (many more stars are available
        with a Th detection only), can be derived
        \textit{indepedently} of model calculations.}

\FullConference{10th Symposium on Nuclei in the Cosmos\\
		 July 27 - August 1 2008\\
		 Mackinac Island, Michigan, USA}

\begin{document}

\section{Introduction}

The first stars that formed from the pristine gas left after the Big
Bang were very massive, of the order of 100\,M$_{\odot}$ [1]. After a
very short life time these co-called Population\,III stars exploded as
supernovae, which then provided the first metals to the interstellar
medium. All subsequent generations of stars formed from chemically
enriched material. Metal-poor stars are early Population\,II objects
and belong to the stellar generations that formed from the non-zero
metallicity gas left behind by the first stars. Due to their low
masses ($\sim0.8$\,M$_{\odot}$) they have extremely long lifetimes
that exceed the current age of the Universe of $\sim14$\,Gyr
years. Hence, these stellar ``fossils'' of the early Universe are
still observable today.

In their atmospheres these old objects preserve information about the
chemical composition of their birth cloud. They thus provide
archaeological evidence of the earliest times of the Universe. In
particular, the chemical abundance patterns provide detailed
information about the formation and evolution of the elements and the
involved nucleosynthesis processes. This knowledge is invaluable for
our understanding of the cosmic chemical evolution and the onset of
star- and galaxy formation. Galactic metal-poor stars are the local
equivalent of the high-redshift Universe. They allow us to derive
observational constraints on the nature of the first stars and
supernovae, and on various theoretical works on the early Universe in
general [2].

Focusing on long-lived low-mass metal-poor main-sequence and giant
stars, we are observing stellar chemical abundances that reflect the
composition of the interstellar medium during their star formation
processes [3]. Stars spend $\sim90\%$ of their lifetime on
the main sequence before they evolve to become red
giants. Main-sequence stars only have a shallow convection zone that
preserves the stars' birth composition over billions of years. Stars
on the red giant branch have deeper convection zones that lead to a
successive mixing of the surface layers with nuclear burning products
from the stellar interior. In the lesser evolved giants the surface
composition has not yet been significantly altered by any such mixing
processes and are useful as tracers of the chemical evolution. Further
evolved stars (e.g., asymptotic giant branch stars) usually have
surface compositions that have been altered through repeated events
that dredge up events of nucleosynthetic burning products.

The main indicator used to determine stellar metallicity is the iron
abundance, [Fe/H], which is defined as \mbox{[A/B]}$ =
\log_{10}(N_{\rm A}/N_{\rm B})_\star - \log_{10}(N_{\rm A}/N_{\rm
B})_\odot$ for the number N of atoms of elements A and B, and $\odot$
refers to the Sun. With few exceptions, [Fe/H] traces the overall
metallicity of the objects fairly well. This review focuses on
metal-poor stars that have around 1/1,000 of the solar Fe abundances,
and are thus able to probe the earliest epochs of nucleosynthesis
processes.  A detailed summary of the history and the different
``classes'' of metal-poor stars and their role in the early Universe
can be found in ref. [4].

\section{Observing the r-Process Signature in the Oldest Stars}
\begin{figure}[!t]
 \begin{center}
 \includegraphics[width=12.7cm, clip=]{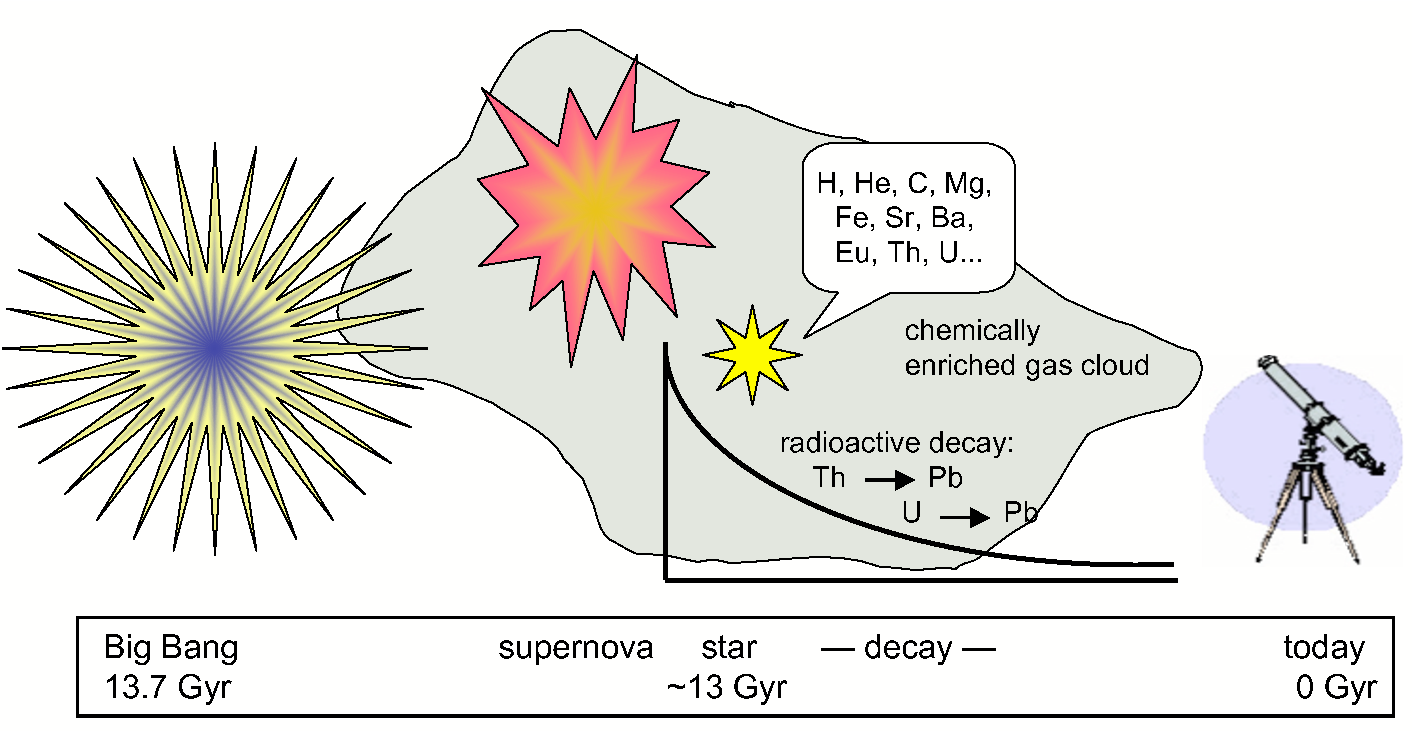}      
   \caption{\label{fingerprint} Schematic view of the formation
   process of r-process-enhanced metal-poor stars. They inherit the
   ``chemical fingerprint'' of a previous-generation supernova explosion.}
 \end{center}
 \end{figure}

All elements except H and He are created in stars during stellar
evolution and supernova explosions. About $5\%$ of metal-poor stars
with $\mbox{[Fe/H]}<-2.5$ contain a strong enhancement of
neutron-capture elements associated with the rapid (r-)
nucleosynthesis process [5] that is responsible for the
production of the heaviest elements in the Universe. In those stars we
can observe the majority (i.e., $\sim70$ of 94) of elements in the
periodic table: the light, $\alpha$, iron-peak, and light and heavy
neutron-capture elements. These elements were not produced in the
observed metal-poor star itself, but in a previous-generation
supernova explosions. The so-called r-process metal-poor stars then
formed from the material that was chemically enriched by such a
supernova.  This is schematically illustrated in
Figure~\ref{fingerprint}. We are thus able to study the ``chemical
fingerprint'' of individual supernova explosions that occurred just
prior to the formation of the observed star. So far, however, the
nucleosynthesis site of the r-process has not yet unambiguously been
identified, but supernovae with progenitor stars of
$8-10$\,M$_{\odot}$ are the most promising locations (see contribution
of Qian in this volume).

The giant HE~1523$-$0901 ($V = 11.1$) was found in a sample of bright
metal-poor stars [6] from the Hamburg/ESO Survey. It
has the so far strongest enhancement in neutron-capture elements
associated with the r-process\footnote{Stars with [r/Fe] $>1.0$; r
represents the average abundance of elements associated with the
r-process.}, $\mbox{[r/Fe]}=1.8$. Its metallicity is
$\mbox{[Fe/H]}=-3.0$ [7]. The spectrum of HE~1523$-$0901
shows numerous strong lines of $\sim25$ neutron-capture elements, such
as those of Sr, Ba, Eu, Os, and Ir. A full discussion of the complete
abundance analysis will be given elsewhere (A.~Frebel et al. 2008, in
preparation). This makes possible a detailed study of the
nucleosynthesis products of the r-process. This fortuitously also
provides the opportunity of bringing together astrophysics and nuclear
physics because these objects act as a ``cosmic lab'' for both fields
of study.

\begin{figure}[!t]
\begin{center}
 \includegraphics[width=10.cm,clip=, bb=30 204 525
 743]{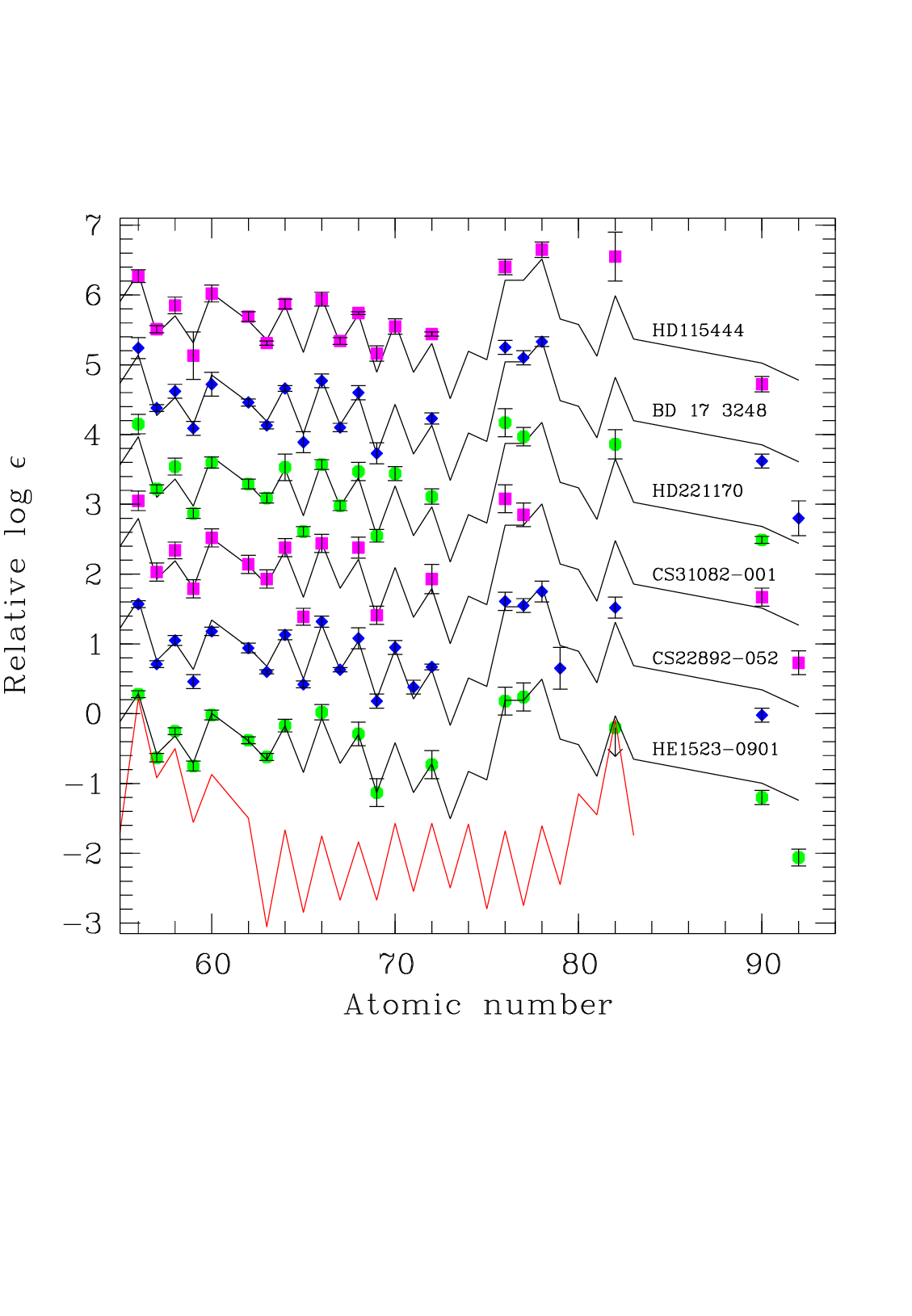}
\caption{\label{pattern} Neutron-capture element abundances ($Z\ge56$)
  of HE~1523$-$0901 and the other most strongly-enhanced r-process
  stars in comparison with those of the scaled solar r-process
  [10]. There is excellent agreement between the stellar data and the
  solar r-process pattern. In red, the scaled solar s-process pattern
  is shown. It does not match the abundances. All patterns are
  arbitrarily offset to allow a visual comparison. References are given
  in Table~1.}
\end{center}
\end{figure}

Although a rarity, HE~1523$-$0901 is not the only star that displays
$\mbox{[r/Fe]}>1.5$. In 1995, the first r-process star was discovered,
CS~22892-052 [8] with $\mbox{[r/Fe]}=1.6$ and in 2001, CS 31082-001
[9] with the same overabundance in these elements.  Their heavy
neutron-capture elements follow the scaled \textit{solar} r-process
pattern, and offered the first vital clues to the universality of the
r-process and the detailed study of the r-process by means of stars.
In Figure~\ref{pattern}, the observed abundance patterns of
HE~1523$-$0901 and other, similar, objects are presented.  As can be
seen, in the mass range $56<Z<77$, the stellar abundances very closely
follow the scaled solar r-process pattern [10].  This repeated
behavior suggests that the r-process is universal -- an important
empirical finding that could not be obtained from any laboratory on
earth. However, there are deviations among the lighter neutron-capture
elements. It is not clear if the neutron-capture abundance patterns
are produced by a single r-process only, or if an additional new
process might need to be invoked in order to explain all
neutron-capture abundances.

 \begin{figure}[!t]
 \begin{center}
  \includegraphics[width=15.cm,clip=, bb=30 224 527
  430]{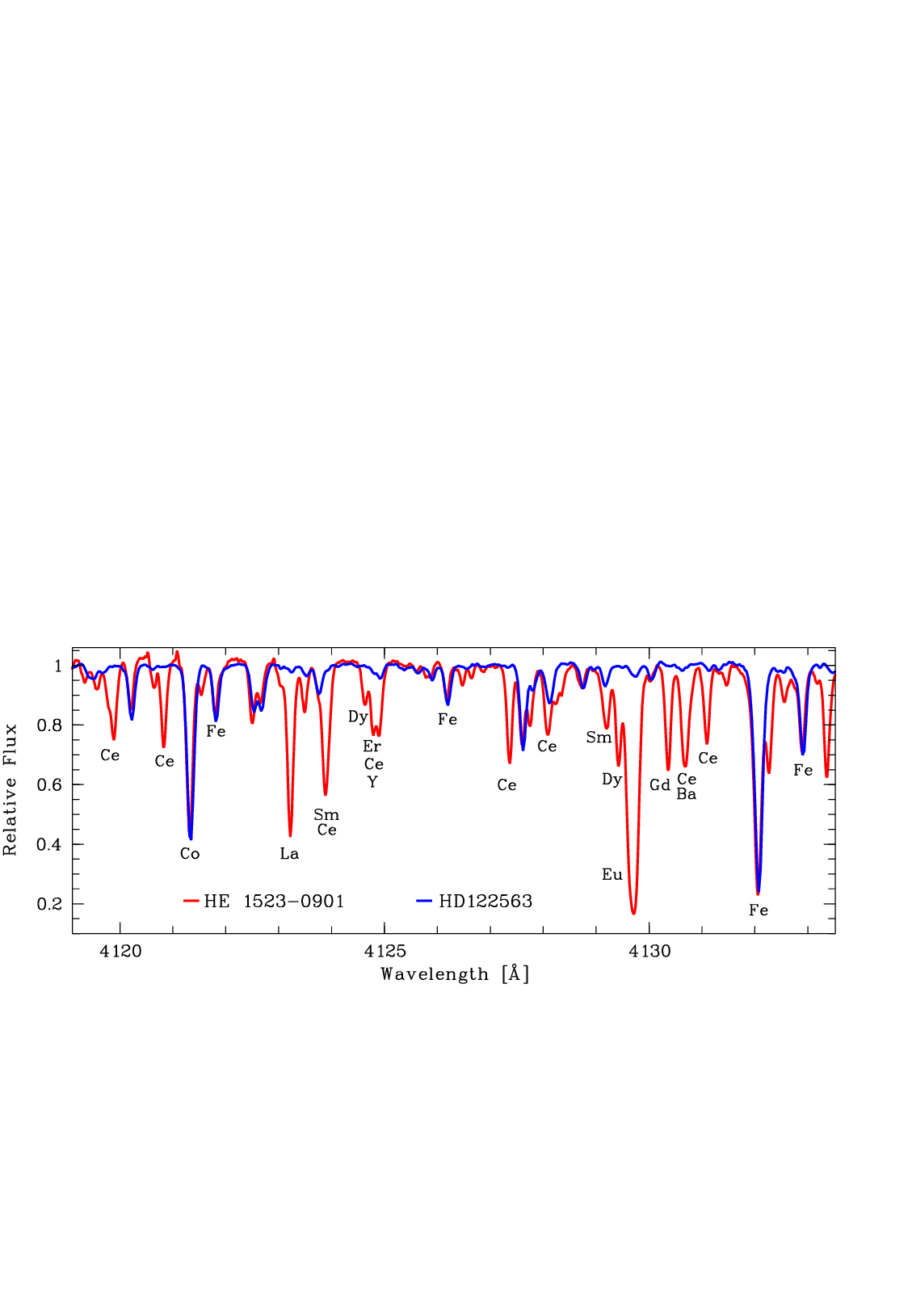}
    \caption{\label{eu_region} Spectral comparison around the Eu\,II
    line at 4129\,{\AA} of the r-process \textit{deficient} star
    HD122563 and the most r-process enhanced star HE~1523$-$0901. Both
    stars have similar temperatures and metallicities.}
 \end{center}
 \end{figure}

In order to find more r-process stars in a systematic fashion, the
Hamburg/ESO R-process Enhance Star survey (HERES) project was carried
out, drawing candidates from the fain Hamburg/ESO survey sample. Over
the course of several years, the HERES project [11] has
indeed led to the discoveries of more r-process-enhanced metal-poor
stars. However, none of them is as strongly enhanced as CS~22892-052,
CS 31082-001, or HE~1523$-$0901, and all of them are relatively faint
($14<B<16$) so that the acquisition of very high $S/N$ data necessary
for the U detection would be somewhat a challenge. HE~1523$-$0901 is
thus the first ``uranium star'' discovered in the Hamburg/ESO Survey,
although drawn from the bright sample (see ref. [6] for a
discussion of these samples). Only a very few r-process enhanced stars
are suitable for a detection of U because the objects need to be
bright, sufficiently cool, strongly overabundant in heavy
neutron-capture elements, and have low C abundances to facilitate the
U detection. It is of great importance, however, to find further
uranium stars. This group of objects (currently with three members)
will provide crucial observational constraints to the study of the
r-process and its possible production site(s).

From the large HERES sample ($\sim350$ stars) it became clear that the
production of neutron-capture elements is decoupled from that of
other, lighter elements with $Z\le30$. The abundance spread of, e.g,
Eu, at low metallicities is extremely large, while for example, Mg, has
a well-defined correlation ($\mbox{[Mg/Fe]}\sim0.4$ for the vast
majority of stars below $\mbox{[Fe/H]}<-1.5$). The different behavior
can be explained with different production mechanisms and sites for
these groups of elements. Although still within supernova explosions,
the progenitor mass range and associated differences in
nucleosynthesis events and timescales likely plays the most important
role. The increasingly large scatter of Eu towards lower metallicities
also suggests that the production of neutron-capture elements might
have been rather sporadic and not was not nearly as well-established
and significant as the production of the lighter elements.

\section{Nucleo-chronometry}

Among the heaviest elements are the long-lived radioactive isotopes
$^{232}$Th (half-life $14$\,Gyr) and $^{238}$U ($4.5$\,Gyr). While Th
is often detectable in r-process stars, U poses a real challenge
because \textit{only one}, extremely weak, line is available in the
optical spectrum. By comparing the abundances of the radioactive Th
and/or U with those of stable r-process nuclei, such as Eu, stellar
ages can be derived. Through individual age measurements, r-process
objects become vital probes for observational ``near-field''
cosmology. Importantly, it also confirms that metal-poor stars with
similarly low Fe abundances and no excess in neutron-capture elements
are similarly old, and that the commonly made assumption about the low
mass (0.6 to 0.8\,M$_{\odot}$) of these survivors is well justified.

Most suitable for such age measurements are cool metal-poor giants
that exhibit such strong overabundances of r-process elements.  Since
CS~22892-052 is very C-rich, however, the U line is blended and not
detectable. Only the Th/Eu ratio could be employed, and an age of
$14$\,Gyr was derived [12].  The U/Th chronometer was first measured
in the giant CS~31082-001 [9] yielding an age of $14$\,Gyr. Since Eu
and Th are much easier to detect than U, the Th/Eu chronometer is then
used to derive stellar ages of r-process metal-poor stars. Compared to
Th/Eu, the Th/U ratio is much more robust to uncertainties in the
theoretically derived production ratio due to the similar atomic
masses of Th and U [13]. Hence, stars displaying Th \textit{and} U are
the most valuable old stars.

\begin{figure}[!t]
\begin{center}
\includegraphics[width=15.2cm,clip=, bb=31 112 530 315]{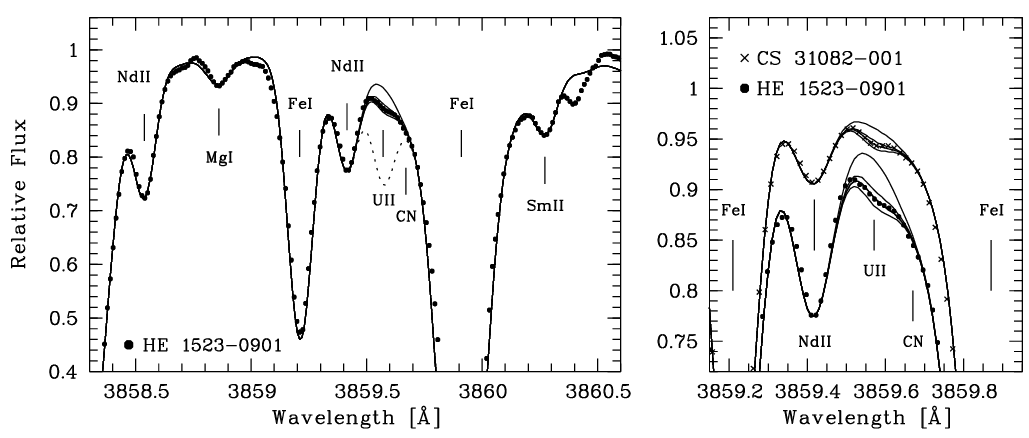}
  \caption{\label{U_region} Spectral region around the U\,II
  line in HE~1523$-$0901 (\textit{filled dots}) and CS~31082-001
  (\textit{crosses}; right panel only). Overplotted are synthetic
  spectra with different U abundances. The dotted line in the left
  panel corresponds to a scaled solar r-process U abundance present in
  the star if no U were decayed. Figure taken from ref. [7].}
\end{center}
\end{figure}

In addition to the heaviest stable elements in HE~1523$-$0901, also
the radioactive isotopes Th and U were measured. In fact, the U
measurement in this star is the currently most reliable one of the
only \textit{three} stars with such detections. Figure~\ref{U_region}
shows the spectral region around the only available optical U line
from which the U abundance was deduced.  For HE~1523$-$0901, the
availability of both the Th and U opened up the possibility for the
first time to use seven different chronometer pairs consisting of Eu,
Os, Ir, Th, and U. It should be noted, that the other star with a
reliable U abundance, CS~31082-001, suffers from what has been termed
an ``actinide boost'' [14]. Compared with the scaled solar
r-process (i.e., with other stable r-process elements), it contains
too much Th and U. Hence, its Th/Eu ratio yields a negative age. The
origin of this issue has yet to be understood. As a result, however,
it has become clear that this star likely has a different
origin.

There are three types of chronometers that involve the abundances of
Th, U and naturally occurring r-process elements [9].

\begin{itemize}
\item[] \mbox{$\Delta t = 46.7[\log{\rm (Th/r)_{initial}} - {\rm
      \log\epsilon(Th/r)_{now}}]$}
\item[] \mbox{$\Delta t = 14.8[\log{\rm (U/r)_{initial}} - {\rm
      \log\epsilon(U/r)_{now}}]$}
\item[] \mbox{$\Delta t = 21.8[\log{\rm (U/Th)_{initial}} - {\rm
      \log\epsilon(U/Th)_{now}}]$}
\end{itemize}

Here, the subscript ``initial'' refers to the theoretically derived
initial production ratio (PR), while the subscript ``now'' refers to
the observed value of the abundance ratio.

The averaged stellar age of HE~1523$-$0901 derived from seven
chronometers involving combinations of Eu, Os, Ir, Th and U is
$\sim13$\,Gyr. Table~\ref{ages} lists the ages derived from the
various abundance ratios, ``chronometers'', measured in a number of
stars, of which some are also shown in Figure~\ref{pattern}. The
employed initial production ratios can be found in the references
given in the table. Such ages provide a lower limit to the age of the
Galaxy and hence, the Universe which is currently assumed to be
13.7\,Gyr [15]. Realistic age uncertainties range from $\sim2$
to $\sim5$\,Gyr.

From a re-determination of the U/Th ratio in CS~31082-001 [16], a
relative age of this star and HE~1523$-$0901 can be derived.
HE~1523$-$0901 is found to be 1.5\,Gyr younger than CS~31082-001,
which is \textit{independent} of the employed production ratio. This
age difference is based on only a 0.07\,dex difference in the observed
U/Th ratios. Given that the observational uncertainties exceed that
ratio difference, the derived ages of the two stars suggest that they
formed at roughly the same time. This is also reflected in their
almost identical metallicity of $\mbox{[Fe/H]}\sim-3.0$.

\begin{table}[!t]
\caption{\label{ages}Stellar ages derived from different abundance ratios}
\begin{center}
{\small
\begin{tabular}{lllll}
\hline
\noalign{\smallskip}
Star  &Age (Gyr) &Abundance ratio &Ref. \\
\noalign{\smallskip}
\hline
\noalign{\smallskip}

 HD115444     &$15.6\pm4$ & Th/Eu & [17]  \\  
 CS 31082-001 &$14.0\pm2$ &U/Th & [16]      \\  
 BD $+17^{\circ}$ 3248 &$13.8\pm4$ & average of several & [18]\\
 CS 22892-052 &$14.2\pm3$ & average of several & [12] \\  
 HD221170     &$11.7\pm3$ & Th/Eu & [19]  \\  
 HE~1523$-$0901&$13.2\pm2$& average of several & [7]\\ 

\noalign{\smallskip}
\hline
\end{tabular}
}
\end{center}
\end{table}


\section{Reverse Engineering: Using Stars to Calibrate Stellar Ages}

Given the large systematic uncertainties in the initial production
ratios which arise from uncertainties in r-process model calculations,
it is not clear at this point, as to how and when these predictions
maybe be significantly improved.  In the absence of such values, one
can resort to ``eliminate'' the problem by having available a large
sample of strongly r-process enhanced metal-poor stars. In principal,
one can use the old age of those stars to \textit{predict} the initial
production ratio by assuming that the stars are, for example, 13\,Gyr
old. Their low metallicity warrants such an assumption, and 13\,Gyr is
a plausible age for an early Population\,II star. Such predicted
production ratios would then not only provide an empirical calibration
for age determinations of other stars, but also offer strong
observational constraints on any r-process model. Suitable for this
procedure would be stars in which as many as possible chronometer
ratios, including ratios involving U, can be measured to provide a good
``internal'' statistic. Then, at least a handful of stars would be
required to obtain some meaningful statistic on the individual
chronometer ratios as measured in several stars. This way, stars in
which only Th is measured (because they are not r-rich enough or too
C-rich, or too faint to allow acquisition of the required very high
$S/N$ data) can be age-dated, independent of any model
calculations. This is important, because the often employed Th/Eu
ratio is subject to large systematic uncertainties in the
theoretically derived production ratio due to the large separation in
atomic masses of Eu and Th.

So far, only HE~1523$-$0901 offers the possibility to serve as
calibrator. Making use of this technique, the results are presented in
Table~\ref{tab2}. It should be kept in mind, however, that in a sample
of one star, the observational errors have a rather large impact on
the resulting ages of the other stars. Nevertheless, the ages
generally are in a range that appears reasonable for this ``trial''
with only one calibration star. Having a better calibration sample at
hand should greatly improve on the age uncertainties.  For obvious
reasons, it is crucial to discover more bright r-process stars in
which we can measure these abundances with the required high-precision
so that they can be employed as nucleo-chronometric age-calibration
stars. Only then will this technique be successful and provide
self-consistent stellar ages as well as crucial observational tests
for current r-process models.

\begin{table}[!t]
\caption{\label{tab2}Reverse engineering: Using HE~1523$-$0901 to
calibrate ages of other metal-poor stars. }
\begin{center}
{\small
\begin{tabular}{lllll}
\hline
\noalign{\smallskip}
Ratio  & Observed ratios &Initial prod. ratio &Derived ages &Stars\\
&&derived from  & (in Gyr)&\\
&&HE~1523$-$0901 &&\\
\noalign{\smallskip}
\hline
\noalign{\smallskip}
Th/Eu& $-$0.62, $-$0.51 &$-$0.222 &18.6, 13.5  &CS 22892-052, BD $+17^{\circ}$ 3248 \\ 
Th/Eu& $-$0.60, $-$0.60 &$-$0.222 &17.7, 17.7  &HD221170, HD115444 \\ 
Th/Os& $-$1.59, $-$1.63 &$-$1.022 &26.6, 28.5  &CS 22892-052, BD $+17^{\circ}$ 3248 \\ 
Th/Ir& $-$1.47, $-$1.48 &$-$1.082 &18.2, 18.6  &CS 22892-052, BD $+17^{\circ}$ 3248 \\ 
U/Eu & $-$1.33          &$-$0.562 &11.4        & BD $+17^{\circ}$ 3248 \\ 
U/Os & $-$2.45          &$-$1.362 &16.1        & BD $+17^{\circ}$ 3248 \\ 
U/Ir & $-$2.30          &$-$1.422 &13.0        & BD $+17^{\circ}$ 3248\\ 
U/Th & $-$0.82          &$-$0.344 &10.4        &CS 31082-001 \\
\noalign{\smallskip}
\hline
\end{tabular}
}
\end{center}
\end{table}

\section{At the End of Everything: Lead}
We also attempted to measure the Pb line at 4057\,{\AA}, the decay
product of Th and U. However, it could not be detected in the current
spectrum of HE~1523$-$0901 because the $S/N$ is not high enough. The
upper limit of $\log\epsilon(\rm Pb)<-0.2$ can be compared with values
calculated based on the decay of Th and U and all other nuclei with
$A\ge210$ into Pb. Following ref. [20], we determine Pb values based
on the decay of $^{238}{\rm U}$ into $^{206}{\rm Pb}$, $^{232}{\rm U}$
into $^{208}{\rm Pb}$, and $^{235}{\rm U}$ into $^{207}{\rm Pb}$,
whereby the last one is based on a theoretically derived ratio of
$^{235}{\rm U}/^{238}{\rm U}$. The total abundance of these three
decay channels amounts to  $\log\epsilon(\rm Pb)=-0.72$. This is in
agreement with our upper limit of $\log\epsilon(\rm Pb)<-0.2$.  There are,
however, also other decay channels through which Pb is built up. Based
on r-process model calculations predictions can be derived for the
total Pb to be measured in HE~1523$-$0901. Preliminary calculations
indicate that this value may be high enough so that with new, higher
$S/N$ data ($S/N$ of 500 or more at 4050\,{\AA}) a detection of this
very weak line should become feasible, or at least provide a much
tighter and more constraining upper limit.

\section{Outlook}

Old metal-poor stars in our Galaxy have been shown to provide crucial
observational clues to the nature of neutron-capture processes, in
particular the r-process. However, even after dedicated searches, only
about two dozens of these stars are known, and only three with any
detection of U. Clearly, more such objects are needed to arrive at
statistically meaningful abundances constraints for various r-process
calculations. At the University of Texas, we have recently started a
new observing project to identify the chemical nature of candidate
metal-poor stars. The ``Chemical Abundances of Stars in the Halo''
(CASH) project is carried out with the Hobby-Eberly Telescope at the
McDonald Observatory in West Texas [21]. With this new program, we aim to
observe up to 1000 metal-poor halo stars over several years to build
up the largest high-resolution spectroscopic database of these
precious objects. It is expected that many ``chemically peculiar''
metal-poor stars (i.e., stars with deviations from the solar ratios)
will be found, including new r-process stars. With an increasing
sample size of these stars, the nucleosynthesis processes of the early
Galaxy can be understood in more and more detail. This is an important
step towards unraveling the chemical nature of the Milky
Way. Combining the chemical abundances with ages of old halo stars, as
well as their kinematic information and the theoretical understanding
of nucleosynthesis \& star formation processes in the early Universe
will finally provide us with new, exciting insight into the entire
formation history of our Galaxy.
\\\\
The author thanks the conference organizers for such a wonderful
meeting. She acknowledges support through the W.~J.~McDonald
Fellowship of the McDonald Observatory.

\end{document}